# DLIMD: Dictionary Learning based Image-domain Material Decomposition for spectral CT

Weiwen Wu, Haijun Yu, Peijun Chen, Fulin Luo, Fenglin Liu, *Member, IEEE*, Qian Wang, Yining Zhu, *Member, IEEE*, Yanbo Zhang, *Senior Member, IEEE,* Jian Feng, Hengyong Yu, *Senior Member, IEEE*

*Abstract*—The potential huge advantage of spectral computed tomography (CT) is its capability to provide accuracy material identification and quantitative tissue information. This can benefit clinical applications, such as brain angiography, early tumor recognition, *etc*. To achieve more accurate material components with higher material image quality, we develop a dictionary learning based image-domain material decomposition (DLIMD) for spectral CT in this paper. First, we reconstruct spectral CT image from projections and calculate material coefficients matrix by selecting uniform regions of basis materials from image reconstruction results. Second, we employ the direct inversion (DI) method to obtain initial material decomposition results, and a set of image patches are extracted from the mode-1 unfolding of normalized material image tensor to train a united dictionary by the K-SVD technique. Third, the trained dictionary is employed to explore the similarities from decomposed material images by constructing the DLIMD model. Fourth, more constraints (i.e., volume conservation and the bounds of each pixel within material maps) are further integrated into the model to improve the accuracy of material decomposition. Finally, both physical phantom and preclinical experiments are employed to evaluate the performance of the proposed DLIMD in material decomposition accuracy, material image edge preservation and feature recovery.

*Index Terms*—spectral computed tomography (CT), material decomposition, image domain, dictionary learning.

## I. Introduction

THE conventional computed tomography (CT) treats x-ray spectrum as single-energy, and an energy integration detector (EID) is employed to collect photons from the spectrum. As a result, different material components may have the same or similar linear attenuation coefficients in the reconstructed CT images. This implies that it fails to discriminate and decompose material components within the imaging objects [1, 2], and it further loses the functional information for CT images. Fortunately, due to the advantages of dose reduction, tissue contrast improvement, quantitative tissue analysis, beam hardening artifacts reduction, material discrimination and decomposition[3], the spectral CT has been developed as an alternative solution. Compared with the traditional CT schemes, spectral CT systems can obtain two or more energy bin projections for the scanned objects. From this view point, the dual energy CT (DECT), as a simple commercialized architecture, has achieved a great success in practical applications [4, 5]. However, the projections from DECT only contain two energy-spectrum measurements. This limits the material decomposition capability of DECT in two basis materials without additional constraints [6, 7].

The development of photon-counting detectors (PCDs) makes spectral CT a hot topic in recent years [8, 9]. A typical spectral CT system equipped with PCD can collect multiple projections for different energy bins within one scan. However, different materials may have similar attenuation coefficients in some of the energy bins. Besides, how to perform a quantitative analysis for tissue distribution is still an open problem. To realize the aims of material discrimination and tissue quantitative analysis, it is necessary to develop advanced material decomposition methods for spectral CT.

Regarding the material decomposition methods for spectral CT, they can be divided into two categories: direct and indirect methods [10]. Specifically, a direct material decomposition method can directly obtain material components from projections [9, 11-13] with known x-ray spectrum. However, it is difficult to achieve x-ray transmission spectral model in practice. Besides, although some regularization prior (for examples, total variation (TV) [9], nonlocal TV [11] and so on) have been considered in such a material decomposition model, the results from direct material decomposition are still sensitive to noise.

Indirect material decomposition methods can be further divided into projection-based and image-based methods [14]. For the projection-based methods, the projections are first decomposed to the specified material sinograms. Then, an image reconstruction algorithm (for example, FBP [15]) is followed. For such material decomposition methods, the errors within the specified basis material sinograms can be magnified in the material maps, and this further comprise material decomposition accuracy. For the image-based methods, the 1st step is to reconstruct spectral CT images from projections, and the 2nd step is to reconstruct material maps. In terms of the 1st step, there are a lot of iterative image reconstruction methods

This work was supported in part by the National Natural Science Foundation of China (No. 61471070 and No. 61501310) and NIH/NIBIB U01 grant (EB017140).

W. Wu, H.J. Yu, P. Chen, F. Liu and J. Feng are with Key Lab of Optoelectronic Technology and Systems, Ministry of Education, Chongqing University, Chongqing 400044, China. F. Luo is with State Key Laboratory of Information Engineering in Surveying, Mapping and Remote Sensing, Wuhan University, Wuhan 430079, China. Q. Wang and H. Yu are with the Department of Electrical and Computer Engineering, University of Massachusetts Lowell, Lowell, MA 01854, USA. Y. Zhu is with School of Mathematical Sciences, Capital Normal University, Beijing, 100048, China. He is also with Beijing Higher Institution Engineering Research Center of Testing and Imaging, Beijing, 100048, China. Y. Zhang is with the PingAn Technology, US Research Lab, Palo Alto, CA 94306, USA.

F. Liu and H. Yu serve as the correspondence authors (e-mail: liufl@cqu.edu.cn and hengyong-yu@ieee.org).

available, such as channel-independent total variation (TV) [16], HighlY constrained backPRojection (HYPR) [17], tight frame sparsity [18], patch-based low-rank [3], spectral prior image constraint compressed sensing (spectral PICCS) [19], prior rank and sparsity [20], tensor dictionary learning (TDL) [21] and its advanced version (L0TDL) [22], nonlocal sparse matrix decomposition [23], spatial-spectral cube matching frame (SSCMF) [24], and non-local low-rank cube-based tensor factorization (NLCTF) [25]. Regarding the $2^{nd}$ step, numerous methods were proposed for DECT[26-29]. However, the methods for multiple-bin spectral CT are relative scarce. For examples, Tao *et.al* [30] developed a prior knowledge aware iterative denoising material decomposition (MD-PKAD) model by considering the Prior Image Constrained Compressed Sensing (PICCS) [31]. Xie *et.al* proposed a multiple constraint model for image domain material decomposition and validated it only by numerical simulation [32]. Besides, the deep learning based multi-material decomposition models [33, 34] are also studied. In this work, we will focus on the $2^{nd}$ step for image domain material decomposition.

The dictionary learning-based methods have obtained a great achievement in the CT field, such as low-dose reconstruction [35], sparse-view reconstruction [36], medical image denoising [37], spectral CT reconstruction [38], *etc*. Indeed, the dictionary learning-based methods can recover image details and features by training a large number of image patches. Similar to the hyperspectral image[39], considering the correlation among different energy channels (i.e., image structure similarities of the same object), the tensor based dictionary learning was developed for spectral CT reconstruction and then obtain excellent performance in our previous studies [21, 22]. As for the multiple material decomposition of spectral CT, multiple material images can be considered as a tensor, leading to a natural idea to establish tensor based dictionary learning for image-domain material decomposition. However, since it is difficult to ensure a rank-1 tensor within a local region for multiple materials, it is inappropriate to adopt the tensor dictionary learning to formulate material decomposition model. Another idea is to develop a multiple-dictionary based decomposition method, i.e., each material corresponds to one dictionary. However, the correlation of material images will be ignored. Besides, because some materials only have simple structures (see iodine contrast agent in Fig. 1(d) in section III), the trained dictionary cannot code material information in the decomposition process. To overcome the aforementioned limitations, we formulate a unified dictionary by training image patches from mode-1 unfolding of normalized material image tensors. Then, we construct a dictionary learning based image-domain material decomposition (DLIMD) model to fully code the sparsity and similarities of material images simultaneously. Finally, the volume conservation and bound of each pixel are introduced into DLIMD model to further improve material decomposition accuracy. The advantages of DLIMD method include good capabilities of noise suppression and edge preservation than other competing methods.

The rest of this paper is organized as follows. In section II, the material decomposition model for spectral CT will be given. In section III, we elaborate the mathematical model and solution for the DLIMD. In section IV, two real datasets are employed to evaluate the proposed decomposition method. In section V, we discuss some related issues and make a conclusion.

## II. MATERIAL DECOMPOSITION MODEL FOR SPECTRAL CT

### *A. Spectral CT imaging model*

Assuming the detected photon number for the $\ell^{th}$ x-ray path within the $n^{th}$ energy window $E_n$ is $y_{n\ell}$ ($1 \leq n \leq N$). It can be modelled as

$$y_{n\ell} = \int_{E_n} I_{n\ell}(E) e^{\int_{r \in \ell} -\sum_{m=1}^M x_m(E,\mathbf{r}) d\mathbf{r}} dE \ , \quad (1)$$

where $\int_{E_n} dE$ integrates over the range of $n^{th}$ energy channel and $\int_{r \in \ell} d\mathbf{r}$ indicates integral along the $\ell^{th}$ x-ray path. In this study, $M$ basis materials within the object is assumed, $x_m(E, \mathbf{r})$ represents the $m^{th}$ ($1 \leq m \leq M$) material linear attenuation coefficient for energy $E$ at position $\mathbf{r}$, and $\sum_{m=1}^M x_m(E, \mathbf{r})$ corresponds to the summation of material attenuation at position $\mathbf{r}$. $I_{n\ell}(E)$ represents the original x-ray photon intensity emitting from the x-ray source for energy $E$.

Considering the basis material expansion, the summation of x-ray attenuation coefficient $\sum_{m=1}^M x_m(E, \mathbf{r})$ in Eq. (1) can be furtherly written as a low-dimensional expansion. That is,

$$\sum_{m=1}^M x_m(E, \mathbf{r}) = \sum_{m=1}^M \vartheta_m(E) f_m(\mathbf{r}), \quad (2)$$

where $\vartheta_m(E)$ represents the mass-attenuation coefficient for the $m^{th}$ material at energy $E$ which can refer to the tables in [40], and $f_m(\mathbf{r})$ is component for $m^{th}$ material at location $\mathbf{r}$. The task of material decomposition is to recover all the material component maps $f_m(\mathbf{r})(1 \leq m \leq M)$ from Eq.(1). If we directly reconstruct material maps from Eq. (1), it can be considered as direct material decomposition [10]. In this study, we only consider the indirect material decomposition methods, i.e., image domain material decomposition, where the first step is to reconstruct the summation of equivalent mass-attenuation coefficient for $\sum_{m=1}^M x_m(\overline{E_n}, \mathbf{r})$. Let $I_{n\ell}^{(0)}$ be the original x-ray photon flux which can be expressed as

$$I_{n\ell}^{(0)} = \int_{E_n} I_{n\ell}(E) dE. \quad (3)$$

Considering the equivalent mass-attenuation coefficient $\sum_{m=1}^M x_m(\overline{E_n}, \mathbf{r})$ and substituting Eqs. (2) and (3) into Eq. (1), we have

$$y_{n\ell} \approx I_{n\ell}^{(0)} e^{\int_{r \in \ell} -\sum_{m=1}^M x_m(\overline{E_n}, \mathbf{r}) d\mathbf{r}}. \quad (4)$$

Considering the multi-energy CT imaging model, the averaged attenuation coefficients $\overline{x_n}(\overline{E_n}, \mathbf{r})$ is employed to replace $\sum_{m=1}^M x_m(\overline{E_n}, \mathbf{r})$. Eq. (4) can be discretized as following

$$y_{n\ell}/I_{n\ell}^{(0)} \approx e^{-\mathbf{A}_{\ell\#} \overline{x_n}(\overline{E_n})}, \quad (5)$$

where $\mathbf{A} \in \mathcal{R}^{L \times J} (J = J_1 \times J_2)$ is the projection matrix, $J_1$ and $J_2$ represent width and height of the averaged attenuation coefficients image, $L$ is the number of total x-ray paths, and $\mathbf{A}_{\ell\#}$ presents the $\ell^{th}$ row of $\mathbf{A}$. $\overline{x_n}(\overline{E_n}) \in \mathcal{R}^J$ is the vectorization of averaged attenuation coefficient image at $n^{th}$ energy window. Then, a logarithm operation is operated on both sides of Eq. (5). By ignoring the index $\ell$, Eq. (5) can be simplified as

$$\boldsymbol{p}_n = -\ln\left(\frac{y_n}{I_n^{(0)}}\right) \approx \mathbf{A}\overline{\boldsymbol{x}_n}. \quad (6)$$

For the image-domain material decomposition method, we can reconstruct multi-energy CT images $\overline{x_n}(n = 1, \ldots, N)$ from the corresponding $\boldsymbol{p}_n(n = 1, \ldots, N)$. There are numerous advanced multi-energy CT image reconstruction techniques, including TDL[21], L$_0$TDL[22], NLCTF [41] and so on. In this study, we focus the image-domain material decomposition, and FBP and NLCTF are employed to obtain $\{\overline{x_n}\}_{n=1}^N$.

### B. Image-domain Material decomposition model

In order to obtain the material component maps $f_m(\mathbf{r})(1 \leq m \leq M)$ from $\{\overline{x_n}\}_{n=1}^N$, Eq.(2) can be further read as a matrix form

$$\begin{bmatrix} \vartheta_{11} & \cdots & \vartheta_{1M} \\ \vdots & \ddots & \vdots \\ \vartheta_{N1} & \cdots & \vartheta_{NM} \end{bmatrix} \begin{bmatrix} f_1 \\ \vdots \\ f_M \end{bmatrix} = \begin{bmatrix} \overline{x_1} \\ \vdots \\ \overline{x_N} \end{bmatrix}, \quad (7)$$

where $\vartheta_{nm}$ represents averaged attenuation coefficients of the $m^{th}$ material at $n^{th}$ energy window, which can be determined by calculating the averaged mass coefficients[42]. Eq. (7) is equivalent to

$$\boldsymbol{\vartheta}\boldsymbol{\mathcal{F}}_{(3)} = \boldsymbol{\mathcal{X}}_{(3)}, \quad (8)$$

where $= \begin{bmatrix} \vartheta_{11} & \cdots & \vartheta_{1M} \\ \vdots & \ddots & \vdots \\ \vartheta_{N1} & \cdots & \vartheta_{NM} \end{bmatrix} \in \mathcal{R}^{N \times M}$, $\boldsymbol{\mathcal{F}} \in \mathcal{R}^{J_1 \times J_2 \times M}$ and $\boldsymbol{\mathcal{X}} \in \mathcal{R}^{J_1 \times J_2 \times N}$ are two tensors which respectively represent the reconstructed images and material images. $\boldsymbol{\mathcal{F}}_{(3)} \in \mathcal{R}^{M \times J}$ and $\boldsymbol{\mathcal{X}}_{(3)} \in \mathcal{R}^{N \times J}$ are the mode-3 unfolding of $\boldsymbol{\mathcal{F}}$ and $\boldsymbol{\mathcal{X}}$.

However, the reconstructed spectral CT images usually contain noise, which can comprise the decomposed material images quality. Considering noise in the reconstructed images, Eq. (8) can be modified as

$$\boldsymbol{\vartheta}\boldsymbol{\mathcal{F}}_{(3)} = \boldsymbol{\mathcal{X}}_{(3)} + \boldsymbol{\epsilon}. \quad (9)$$

To recover spectral CT images from Eq. (9), we have the following least square linear program problem:

$$\min_{\boldsymbol{\mathcal{F}}} \frac{1}{2}\|\boldsymbol{\mathcal{X}}_{(3)} - \boldsymbol{\vartheta}\boldsymbol{\mathcal{F}}_{(3)}\|_F^2, \quad (10)$$

where $\|\cdot\|_F$ is Frobenius norm. Eq. (10) can be viewed as image domain basis material decomposition model. The simple direct inversion (DI) method [26] can be employed to solve Eq.(10) and obtain:

$$\boldsymbol{\mathcal{F}}_{(3)} = (\boldsymbol{\vartheta}^T\boldsymbol{\vartheta})^{-1}\boldsymbol{\vartheta}^T\boldsymbol{\mathcal{X}}_{(3)}. \quad (11)$$

To improve the decomposed image quality, using regularization is a feasible strategy to constrain the solution during iteration process, and Eq. (10) can be reformulated as

$$\min_{\boldsymbol{\mathcal{F}}} \left(\frac{1}{2}\|\boldsymbol{\mathcal{X}}_{(3)} - \boldsymbol{\vartheta}\boldsymbol{\mathcal{F}}_{(3)}\|_F^2 + \lambda R(\boldsymbol{\mathcal{F}})\right), \quad (12)$$

where $\lambda$ is a regularization factor to balance the data fidelity term $\frac{1}{2}\|\boldsymbol{\mathcal{X}}_{(3)} - \boldsymbol{\vartheta}\boldsymbol{\mathcal{F}}_{(3)}\|_F^2$ and regularization term $R(\boldsymbol{\mathcal{F}})$.

## III. DICTIONARY LEARNING BASED IMAGE-DOMAIN MATERIAL DECOMPOSITION (DLIMD)

### A. Dictionary learning

A set of image patches $\mathbf{x}_i \in \mathcal{R}^{s \times s}$, $i = 1, \ldots, I$, are extracted from the training datasets and unfolded as vectors to train the dictionary $\mathbf{D} \in \mathcal{R}^{S \times T}$, where $S = s \times s$ and $T$ represents the number of atoms. The goal of dictionary learning is to find sparse representation coefficients $\boldsymbol{\alpha} \in \mathcal{R}^{T \times I}$ using the dictionary $\mathbf{D}$. This can be established by solving the following optimization problem:

$$\min_{\mathbf{D},\boldsymbol{\alpha}} \frac{1}{2}\sum_{i=1}^{I}\|\mathbf{x}_i - \mathbf{D}\boldsymbol{\alpha}_i\|_F^2 \quad \text{s.t.} \ \|\boldsymbol{\alpha}_i\|_0 \leq L_0, \quad (13)$$

where $L_0$ is the sparsity level, $\|\cdot\|_0$ is the quasi-$l_0$ norm, $\boldsymbol{\alpha}_i \in \mathcal{R}^{T \times 1}$ is sparse representation coefficients for $i^{th}$ image patch. Eq. (13) is a constrained linear programming problem, which can be converted into the following unconstrained problem:

$$\min_{\mathbf{D},\boldsymbol{\alpha}} \frac{1}{2}\sum_{i=1}^{I}(\|\mathbf{x}_i - \mathbf{D}\boldsymbol{\alpha}_i\|_F^2 + v_i\|\boldsymbol{\alpha}_i\|_0), \quad (14)$$

where $v_i$ is a Lagrange multiplier. Eq. (14) can be solved by utilizing an alternating minimization scheme. The first step is to update $\boldsymbol{\alpha}_i$ by fixing the dictionary $\mathbf{D}$,

$$\min_{\boldsymbol{\alpha}} \frac{1}{2}\sum_{i=1}^{I}(\|\mathbf{x}_i - \mathbf{D}\boldsymbol{\alpha}_i\|_F^2 + v_i\|\boldsymbol{\alpha}_i\|_0). \quad (15)$$

Eq.(15) can be optimized using the matching pursuit (MP) [43] or orthogonal matching pursuit (OMP) algorithm [44]. The second step is to update the dictionary with a fixed sparse representation coefficients $\boldsymbol{\alpha}$. There are many methods available to train the dictionary, such as K-SVD [45] and online learning technique[46].

### B. DLIMD Method

Considering the general image domain material decomposition model Eq. (12), similar to the dictionary learning for low-dose CT reconstruction [35], the proposed Dictionary Learning based Image-domain Material Decomposition (DLIMD) for spectral CT can be constructed as follow:

$$\min_{\boldsymbol{\mathcal{F}},\{\boldsymbol{\beta}_m\}_{m=1}^M} \left(\begin{array}{c} \frac{1}{2}\|\boldsymbol{\mathcal{X}}_{(3)} - \boldsymbol{\vartheta}\boldsymbol{\mathcal{F}}_{(3)}\|_F^2 \\ + \sum_{m=1}^{M}\frac{\lambda_m}{2}\sum_{i=1}^{I}\left(\|\mathcal{H}_i(F_m) - \widehat{\mathbf{D}}\boldsymbol{\beta}_{mi}\|_F^2 \\ +v_{mi}\|\boldsymbol{\beta}_{mi}\|_0 \end{array}\right)\right), \quad (16)$$

where $F_m$ is $m^{th}$ channel of material images $\boldsymbol{\mathcal{F}}$, $\boldsymbol{\beta}_{mi} \in \mathcal{R}^{T \times 1}$ is sparse representation coefficients for $i^{th}$ image patch from $m^{th}$ material image and $\boldsymbol{\beta}_m = \{\boldsymbol{\beta}_{mi}\}_{i=1}^{I}$. $\mathcal{H}_i(F_m)$ is the $i^{th}$ image patch extraction operator from $F_m$. $\widehat{\mathbf{D}}$ is the trained dictionary. In this study, we use a united dictionary for different material decomposition. High-quality dictionary is beneficial for sparse representation and accuracy of material decomposition. A natural idea is to train different $\widehat{\mathbf{D}}$ for different materials. However, this strategy will be limited by three reasons. First, a specific material map may only contain a few image features, i.e., bone and iodine contrast in Fig.1. It is difficult for the trained $\widehat{\mathbf{D}}$ to encode the image features and reduce sparse representation level during the process of material decomposition, and this comprises the material decomposition accuracy. Second, training different $\widehat{\mathbf{D}}$ is time consuming, which implies higher computational cost is needed with various materials in practice. Third, if we train the dictionary $\widehat{\mathbf{D}}$ for each material, the correlation between different material maps will be lost. To overcome these limitations, we formulate a unified dictionary $\widehat{\mathbf{D}} \in \mathcal{R}^{S \times T}$ in this study. For a given spectral CT dataset, we can obtain the FBP or other reconstruction results and apply DI method to achieve material decomposition results $\mathcal{F}$. Then, before training the dictionary $\widehat{\mathbf{D}}$, we should normalize material images to avoid data inconsistency of the pixel values from different materials. After that, a set of image patches are extracted from $\mathcal{F}'_{(1)} \in \mathcal{R}^{J_1 \times (J_2 \times M)}$, the mode-$1$ unfolding of normalized image $\mathcal{F}'$, to train a global dictionary $\widehat{\mathbf{D}}$. Although the DI material maps contain noise, our following experiments demonstrate that the dictionary training can preserve image information and against noise well. Finally, the dictionary $\widehat{\mathbf{D}}$ can be learned using the K-SVD algorithm [45]. The advantages of the formulated unified dictionary $\widehat{\mathbf{D}}$ are threefold. First, it can fully encode similarities within different materials images (for examples, the image structures within blue boxes from bone and soft tissue, the image structures within red boxes from soft tissue and iodine contrast in Fig. 1) during the material decomposition process. Second, it can enhance the redundancy with the trained dictionary $\widehat{\mathbf{D}}$. This is good to encode current material image structures by employing the atoms from other material image to enhance sparse representation. Third, we can also save training time to some extent in practice.

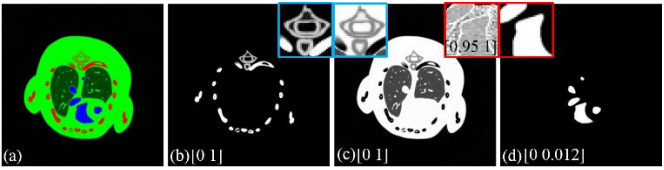

Fig.1 Numerical simulation mouse phantom used in [21] consists of three materials, i.e. bone, soft tissue and iodine. (a), (b), (c) and (d) represent the phantom, bone, soft tissue and iodine, respectively.

To further improve the accuracy of material decomposition, more constraints should be considered. For example, it is reasonable to assume that bone and contrast agent don't co-exist in a given pixel, and this assumption is employed in our experiments. If the air is also treated as one basis material, the summation of pixel values at the same location in different material images should be equal to one [7], i.e.,

$$\left(\sum_{m=1}^{M} \mathcal{F}_{j_1 j_2 m}\right) + AIR_{j_1 j_2} = 1 \ (1 \leq j_1 \leq J_1, 1 \leq j_2 \leq J_2), \quad (17)$$

where $\mathcal{F}_{j_1 j_2 m}$ represents the pixel value at $(j_1, j_2, m)^{th}$ location. $AIR$ is the air map and $AIR_{j_1 j_2}$ represents the pixel value at $(j_1, j_2)^{th}$ location. Besides, the pixel value within $\mathcal{F}$ should be in the range of [0 1], i.e.,

$$0 \leq \mathcal{F}_{j_1 j_2 m} \leq 1. \quad (18)$$

To further constrain the solution for optimization and improve the accuracy of material decomposition, Eqs. (17)-(18) can be treated as two conditions and are incorporated into the material decomposition model Eq. (16). Then, we have:

$$\min_{\mathcal{F}, \{\boldsymbol{\beta}_m\}_{m=1}^{M}} \left( \begin{array}{c} \frac{1}{2}\|\boldsymbol{\mathcal{X}}_{(3)} - \boldsymbol{\vartheta}\mathcal{F}_{(3)}\|_F^2 + \\ \sum_{m=1}^{M} \frac{\lambda_m}{2} \sum_{i=1}^{I} \left(\|\mathcal{H}_i(F_m) - \widehat{\mathbf{D}}\boldsymbol{\beta}_{mi}\|_F^2 + v_{mi}\|\boldsymbol{\beta}_{mi}\|_0 \right) \end{array} \right)$$
$$s.t. \left(\sum_{m=1}^{M} \mathcal{F}_{j_1 j_2 m}\right) + AIR_{j_1 j_2} = 1,$$
$$0 \leq \mathcal{F} \leq 1, \forall j_1, j_2. \quad (19)$$

To solve Eq. (19), we first introduce a tensor $\boldsymbol{\mathcal{U}}$ to replace $\mathcal{F}$, and Eq. (19) can be converted into the following constrained programmable problem:

$$\min_{\mathcal{F}, \{\boldsymbol{\beta}_m\}_{m=1}^{M}, \boldsymbol{\mathcal{U}}} \left( \begin{array}{c} \frac{1}{2}\|\boldsymbol{\mathcal{X}}_{(3)} - \boldsymbol{\vartheta}\mathcal{F}_{(3)}\|_F^2 + \\ \sum_{m=1}^{M} \frac{\lambda_m}{2} \sum_{i=1}^{I} \left(\|\mathcal{H}_i(\mathbf{U}_m) - \widehat{\mathbf{D}}\boldsymbol{\beta}_{mi}\|_F^2 + v_{mi}\|\boldsymbol{\beta}_{mi}\|_0 \right) \end{array} \right)$$
$$s.t. \left(\sum_{m=1}^{M} \mathcal{F}_{j_1 j_2 m}\right) + AIR_{j_1 j_2} = 1, \boldsymbol{\mathcal{U}} = \mathcal{F},$$
$$0 \leq \mathcal{F} \leq 1, \forall j_1, j_2. \quad (20)$$

Eq. (20) is equal to the following unconstrained optimization problem with certain conditions:

$$\min_{\mathcal{F}, \{\boldsymbol{\beta}_m\}_{m=1}^{M}, \boldsymbol{\mathcal{U}}} \left( \begin{array}{c} \frac{1}{2}\|\boldsymbol{\mathcal{X}}_{(3)} - \boldsymbol{\vartheta}\mathcal{F}_{(3)}\|_F^2 + \frac{\eta}{2}\|\boldsymbol{\mathcal{U}} - \mathcal{F}\|_F^2 \\ \sum_{m=1}^{M} \frac{\lambda_m}{2} \sum_{i=1}^{I} \left(\|\mathcal{H}_i(\mathbf{U}_m) - \widehat{\mathbf{D}}\boldsymbol{\beta}_{mi}\|_F^2 + v_{mi}\|\boldsymbol{\beta}_{mi}\|_0 \right) \end{array} \right)$$
$$s.t. \left(\sum_{m=1}^{M} \mathcal{F}_{j_1 j_2 m}\right) + AIR_{j_1 j_2} = 1,$$
$$0 \leq \mathcal{F} \leq 1, \forall j_1, j_2. \quad (21)$$

Here $\eta > 0$ is an optimization factor to balance the estimation $\boldsymbol{\mathcal{U}}$ for $\mathcal{F}$. The objection function Eq. (21) can be divided into the following two sub-problems:

$$\min_{\mathcal{F}} \left(\frac{1}{2}\|\boldsymbol{\mathcal{X}}_{(3)} - \boldsymbol{\vartheta}\mathcal{F}_{(3)}\|_F^2 + \frac{\eta}{2}\|\mathcal{F} - \boldsymbol{\mathcal{U}}^{(k)}\|_F^2 \right) \quad s.t.$$
$$\left(\sum_{m=1}^{M} \mathcal{F}_{j_1 j_2 m}\right) + AIR_{j_1 j_2} = 1, 0 \leq \mathcal{F} \leq 1, \forall j_1, j_2, \quad (22a)$$

$$\min_{\boldsymbol{\mathcal{U}}, \{\boldsymbol{\beta}_m\}_{m=1}^{M}} \left( \begin{array}{c} \frac{\eta}{2}\|\boldsymbol{\mathcal{U}} - \mathcal{F}^{(k+1)}\|_F^2 \\ + \sum_{m=1}^{M} \frac{\lambda_m}{2} \sum_{i=1}^{I} \left( \begin{array}{c} \|\mathcal{H}_i(\mathbf{U}_m) - \widehat{\mathbf{D}}\boldsymbol{\beta}_{mi}\|_F^2 \\ + v_{mi}\|\boldsymbol{\beta}_{mi}\|_0 \end{array} \right) \end{array} \right), \quad (22b)$$

*i) $\mathcal{F}$ sub-problem:* Regarding the optimization of Eq.(22a), there are two strategies. A natural first strategy is to treat the air as a basis material to $\mathcal{F}$. In this case, the material attenuation matrix $\boldsymbol{\vartheta}$ can be modified as $\boldsymbol{\vartheta}' = \begin{bmatrix} \vartheta_{11} & \cdots & \vartheta_{1M} & \varepsilon \\ \vdots & \ddots & \vdots & \vdots \\ \vartheta_{N1} & \cdots & \vartheta_{NM} & \varepsilon \end{bmatrix} \in \mathcal{R}^{N \times (M+1)}$, where $\varepsilon$ is a small positive value to represent air attenuation coefficient. The air can increase the number of basis materials in this strategy, and it can result in instability of material decomposition and comprise the decomposition accuracy, especially in the case of $M > N$. The second strategy for solving Eq. (22a) can be divided into two steps. The first step is to solve the objection function Eq. (23):

$$\min_{\mathcal{F}} \left( \frac{1}{2} \| \mathcal{X}_{(3)} - \vartheta \mathcal{F}_{(3)} \|_F^2 + \frac{\eta}{2} \| \mathcal{F} - \mathcal{U}^{(k)} \|_F^2 \right)$$

$$s.t. \left( \sum_{m=1}^{M} \mathcal{F}_{j_1 j_2 m} \right) = 1, \forall j_1, j_2, \quad 0 \leq \mathcal{F} \leq 1. \quad (23)$$

Considering Eq. (23) with pixel level, it can be furtherly written as

$$\min_{\mathcal{F}} \frac{1}{2} \sum_{j=1}^{J} \| \mathcal{X}_{(3)\#j} - \vartheta \mathcal{F}_{(3)\#j} \|_F^2 + \frac{\eta}{2} \sum_{j_1=1}^{J_1} \sum_{j_2=1}^{J_2} \| \mathcal{F}_{j_1 j_2 \#} - \mathcal{U}^{(k)}_{j_1 j_2 \#} \|_F^2$$

$$s.t. \left( \sum_{m=1}^{M} \mathcal{F}_{j_1 j_2 m} \right) = 1 \quad \forall j_1, j_2 \quad 0 \leq \mathcal{F}_{j_1 j_2 \#} \leq 1, (24)$$

where $\mathcal{X}_{(3)\#j} = [\overline{x_{1j}}, \ldots, \overline{x_{Nj}}]^T$, $\mathcal{F}_{(3)\#j} = [f_{1j}, \ldots, f_{Mj}]^T$, $\mathcal{F}_{j_1 j_2 \#} = [\mathcal{F}_{j_1 j_2 1}, \ldots, \mathcal{F}_{j_1 j_2 M}]^T$ and $\mathcal{U}^{(k)}_{j_1 j_2 \#} = [\mathcal{U}^{(k)}_{j_1 j_2 1}, \ldots, \mathcal{U}^{(k)}_{j_1 j_2 M}]^T$. Note that $j = j_1 \times j_2$, Eq. (24) is equivalent to

$$\min_{\mathcal{F}} \sum_{j_1=1}^{J_1} \sum_{j_2=1}^{J_2} \left( \frac{1}{2} \| \mathcal{X}_{\#j_1 j_2} - \vartheta \mathcal{F}_{j_1 j_2 \#} \|_F^2 + \frac{\eta}{2} \| \mathcal{F}_{j_1 j_2 \#} - \mathcal{U}^{(k)}_{j_1 j_2 \#} \|_F^2 \right)$$

$$s.t. \left( \sum_{m=1}^{M} \mathcal{F}_{j_1 j_2 m} \right) = 1, \quad 0 \leq \mathcal{F}_{j_1 j_2 \#} \leq 1 \; \forall j_1, j_2, \quad (25)$$

We optimize Eq. (25) by minimizing the following problem:

$$\min_{\mathcal{F}_{j_1 j_2 \#}} \left( \frac{1}{2} \| \mathcal{X}_{\#j_1 j_2} - \vartheta \mathcal{F}_{j_1 j_2 \#} \|_F^2 + \frac{\eta}{2} \| \mathcal{F}_{j_1 j_2 \#} - \mathcal{U}^{(k)}_{j_1 j_2 \#} \|_F^2 \right) \forall j_1, j_2$$

$$s.t. \left( \sum_{m=1}^{M} \mathcal{F}_{j_1 j_2 m} \right) = 1 \quad 0 \leq \mathcal{F}_{j_1 j_2 \#} \leq 1. \quad (26)$$

Eq. (26) is a constrained convex programmable optimization problem. It is equal to the following optimization problem

$$\min_{\mathcal{F}_{j_1 j_2 \#}} \frac{1}{2} \| (\vartheta^T \vartheta + \eta \mathbf{I}) \mathcal{F}_{j_1 j_2 \#} - (\vartheta^T \mathcal{X}_{\#j_1 j_2} + \eta \mathcal{U}^{(k)}_{j_1 j_2 \#}) \|_F^2$$

$$\forall j_1, j_2 \quad s.t. \left( \sum_{m=1}^{M} \mathcal{F}_{j_1 j_2 m} \right) = 1, \quad 0 \leq \mathcal{F}_{j_1 j_2 \#} \leq 1. \quad (27)$$

Eq. (27) is a constrained least square problem, which can be easily solved. The second step is to find the air region by operating a threshold method on the $\mathcal{F}^{(k+1)}$. Again, if the pixel value within $\mathcal{F}^{(k+1)}$ is greater than a given threshold, the pixel value can be replaced by direct inversion. The given threshold is set as 0.99 in this work.

To compare strategies 1 and 2, Fig.2 shows the material decomposition results using direct inversion (DI) method from FBP images with noise-free projections. From Fig. 2, it can be seen that soft tissue results using strategy 1 contain severe strip artifacts. However, the artifacts are suppressed by using strategy 2.

*ii) $\mathcal{U}$ sub-problem:* As for the optimization problem of Eq.(22b), it can be re-written as

$$\min_{\mathcal{U}, \{\beta_m\}_{m=1}^M} \sum_{m=1}^{M} \left( \frac{1}{2} \| \mathbf{U}_m - \mathbf{F}_m^{(k+1)} \|_F^2 + \frac{\tau_m}{2} \times \sum_{i=1}^{I} \left( \| \mathcal{H}_i(\mathbf{U}_m) - \widehat{\mathbf{D}} \beta_{mi} \|_F^2 + v_{mi} \| \beta_{mi} \|_0 \right) \right)$$

.(28)

Eq. (28) can be divided into $M$ sub-problems. Again, $\mathbf{U}_m (m = 1, \ldots, M)$ can be updated independently. Here, Eq. (28) can be furtherly read as

$$\min_{\mathbf{U}_m, \beta_m} \left( \frac{1}{2} \| \mathbf{U}_m - \mathbf{F}_m^{(k+1)} \|_F^2 + \frac{\tau_m}{2} \sum_{i=1}^{I} \left( \| \mathcal{H}_i(\mathbf{U}_m) - \widehat{\mathbf{D}} \beta_{mi}^{(k+1)} \|_F^2 + v_{mi} \| \beta_{mi} \|_0 \right) \right), 1 \leq m \leq M. (29)$$

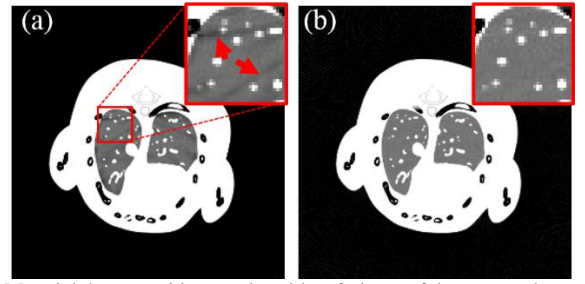

Fig.2 Material decomposition results with soft tissue of the mouse phantom. (a) and (b) represent the results using strategies 1 and 2, respectively.

Eq. (29) is a typical dictionary learning based image denoising model, and it can be solved by the method in [47]. During the process of dictionary learning based image denoising, the parameters of sparsity level $L$ and tolerance of representation error $\varepsilon$ play important roles in controlling the dictionary quality and material decomposition accuracy. In this study, the sparsity level $L$ is set as the same value for different materials. However, the tolerance of representation error $\boldsymbol{\varepsilon} = \varepsilon_m (m = 1, \ldots, M)$ should be careful chosen. The overall pseudo-codes of the DLIMD algorithm is listed in Algorithm I.

| Algorithm I: DLIMD |
|---|
| **Input:** $\eta, \varepsilon, L, T, K$ and other parameters; Initialization of $\mathcal{F}^{(0)} = \mathbf{0}, \mathcal{U}^{(o)} = \mathbf{0}, k = 0$. |
| **Output:** Material decomposition tensor $\mathcal{F}$. |
| **Part I: Dictionary training** |
| 1: Reconstructing spectral CT images using certain image reconstruction method; |
| 2: Formulating the material attenuation matrix using reconstruction results; |
| 3: Decomposing the reconstructed images using DI method; |
| 4: Normalizing the DI results; |
| 5: Extracting image patches to form a dictionary training dataset; |
| 6: Training a dictionary using the K-SVD technique. |
| **Part II: Material decomposition** |
| 7: **While** not convergence **do** |
| 8: Updating $\mathcal{F}^{(k+1)}$ using Eq. (27); |
| 9: Finding and processing the pixels in $\mathcal{F}^{(k+1)}$ using DI technique; |
| 10: Updating $\mathcal{U}$ and $\{\beta_m\}_{m=1}^{M}$ using Eq. (29); |
| 11: $k=k+1$; |
| 12: End **While** |

### C. Algorithm implementation details

From the pseudo-codes of DLIMD method, it can be divided into two main parts: dictionary learning and material decomposition. To implement the dictionary learning, we first reconstruct spectral CT images from projections using analytic/iterative reconstruction methods. In this study, both FBP and NLCTF[25] methods are employed to implement image reconstruction. Calculating the basis material attenuation matrix by selecting uniform regions is a key step for material decomposition. Then, we adopt the DI method to decompose reconstructed images into basis material maps using the formulated material attenuation matrix. Finally, $10^4$ image patches with size of $8 \times 8$ are extracted from normalized material images to train dictionary $\hat{D}$ using the K-SVD algorithm. The dictionary $\hat{D}$ is overcomplete to enforce the sparsity level. Again, the number of atoms in a dictionary should be much greater than the size of an atom, i.e., $T \gg s \times s$. Usually, $T$ should be greater than $4 \times s \times s$. Here, the number of atoms $T$ within the dictionary $\hat{D}$ is set as 512. The sparsity level $L^T$ in the dictionary training can be set empirically from 5 to 10, and it is uniformly set as 6. The iteration number to train dictionary is uniformly set as 200. Fig. 3 shows three trained dictionaries and will be used in the following experiments.

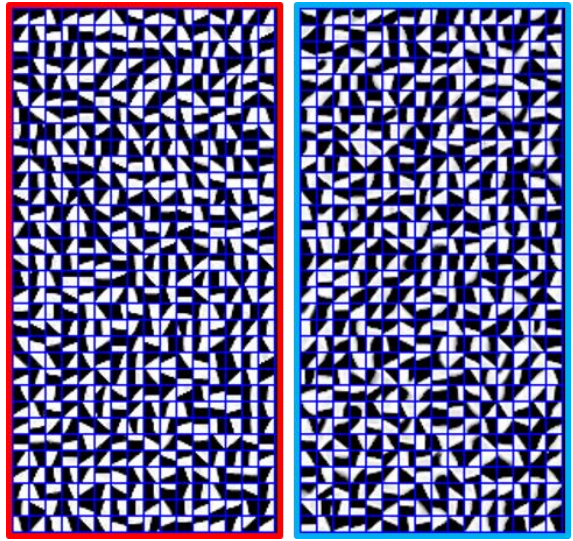

Fig. 3. The dictionaries used in our experiments. 1$^{st}$-2$^{nd}$ columns represent the dictionaries for physical phantom and preclinical experiments.

Especially, Eq. (27) is solved by adopting the *lsqlin* function in Matlab 2014(b). The parameter selections of sparsity representation level $L$ and $\varepsilon = \varepsilon_m (m = 1, ..., M)$ are dependent on specified application. All the methods are stopped after 30 iterations.

## IV. EXPERIMENTS AND RESULTS

To evaluate the performance of our proposed DLIMD method in material decomposition, two real datasets are employed. The advantages are demonstrated by comparing with the DI and total variation-regularization material decomposition (TVMD) methods. To quantitatively evaluate the performance of all material decomposition techniques, the root means square error (RMSE), peak-signal-to-noise ratio (PSNR) and structural similarity (SSIM) are employed. The optimized parameters used in all experiments are listed in Table I.

Table I. Parameters used in the proposed DLIMD method for experiments

| Parameters | $\eta$ | $L$ | $\varepsilon_{Al/bone,Water/Tisuue,Iodine/GNP}$ |
|---|---|---|---|
| Physical phantom | 0.003 | 10 | (0.08, 0.03, 0.004) |
| Preclinical (FBP) | 0.001 | 12 | (0.004, 0.012, 0.05) |
| Preclinical (NLCTF) | 0.001 | 12 | (0.001, 0.004, 0.004) |

### A. Physical Phantom

A physical phantom consists of three basis materials, i.e., water, iodine and aluminum, is employed to evaluate our method first. As shown in Fig. 4, the phantom contains five cylinders, and each of them represents different basis material or different concentrations of iodine solution. Aluminum is used to mimic bone. The spectral CT system is equipped with a micro-focus x-ray source (YXLON, 225Kv) and a flat-panel PCD (Xcounter, XC-Hydra FX20). For the central slice, the PCD contains 2048 detector cells and each of them covers a length of 0.1 mm. Here, every 4 cells are combined to reduce noise. Because the PCD only has two energy channels, the phantom is scanned twice to obtain 4 energy bin projections. The full-scan projections are collected from 1080 uniformly distributed views with 137kV x-ray source. The distances from the x-ray source to rotational center and detector are 182.68 mm and 440.50 mm, respectively. The FOV diameter is calculated as 82.6 mm. Since the reconstructed material image matrix is 256$\times$256, each pixel covers an area of 0.324$\times$0.324 mm$^2$.

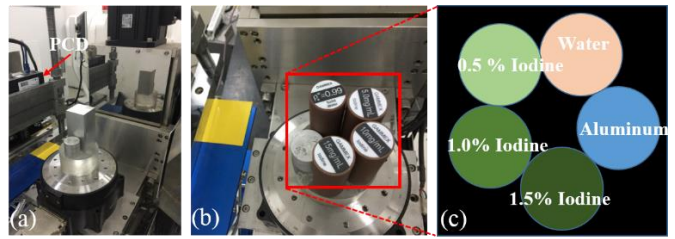

Fig. 4. Setups of physical phantom experiments. (a) is the spectral CT system, (b) and (c) represent the physical phantom.

To evaluate the proposed material decomposition methods in the case of analytic reconstruction results, Fig. 5 shows the reconstructed spectral CT images from four different energy bins using FBP. Because the x-ray has beam hardening, the aluminum cylinder region is slightly damaged by metal artifacts. This can comprise the material decomposition accuracy to some extent. In this study, to validate the advantages of constraint Eq. (17) in material decomposition model, the DI without equivalent transformation constraint (DIWET), i.e., Eq. (17), is also compared. The material decomposition results with three basis materials (aluminum, water and iodine) are given in Fig. 6. From Fig. 6 it can be observed that the constraint Eq. (17) can improve material images quality remarkably compared with the DIWET and DI results. Besides, the material decomposition with regularization terms can also provide higher image quality than that obtained by the DI method. In terms of aluminum results, the proposed DLIMD method can better protect image edges and avoid the blocky artifacts compared with the TVMD method. As for the decomposed water results, the proposed DLIMD can provide much clearer image edge with higher accuracy, which is confirmed by the magnified ROI. Regarding the iodine results, three cylinders from the DLIMD are more complete and uniform than those obtained by other decomposition methods.

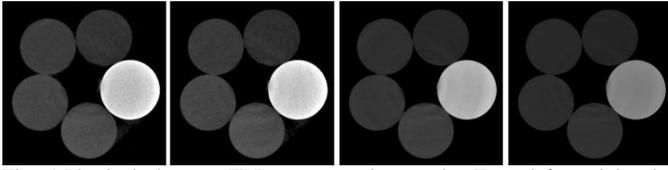

Fig. 5 Physical phantom FBP reconstruction results. From left to right, the images are for 1st -4th energy bins with a display window [0 1.3]cm$^{-1}$.

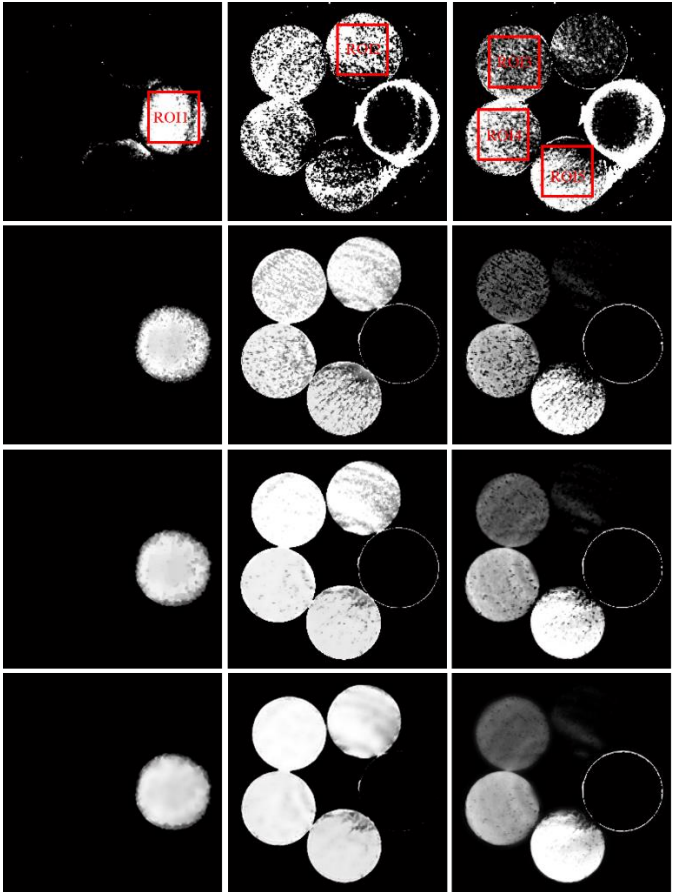

Fig. 6 Material decomposition results from Fig.5. From left to right, the columns represent the decomposition results of aluminum, water and iodine, where the display windows are [0.5 1], [0.8 1] and [0 0.015]. From top to bottom, the rows represent the results decomposed by DIWET, DI, TVMD and DLIMD methods, respectively.

To further evaluate the performance of DLIMD algorithm in improving material decomposition, five region of interests (ROI) indicated with 1-5 are extracted from different material maps and their quantitative evaluation results are listed in Table II. From Table II, we can see the performance of DIWET is always the worst. The proposed method always performs the best followed by the TV-based and DI methods.

Table II. Quantitative evaluation results of ROIs 1-5.

|  |  | DIWET | DI | TVMD | DLIMD |
|---|---|---|---|---|---|
| ROI-1 | RMSE | 0.1890 | 0.0850 | 0.0827 | **0.0804** |
|  | PSNR | 14.471 | 21.406 | 21.649 | **21.893** |
|  | SSIM | 0.9278 | 0.9542 | 0.9823 | **0.9925** |
| ROI-2 | RMSE | 0.2631 | 0.0324 | 0.0318 | **0.0273** |
|  | PSNR | 11.599 | 29.793 | 29.964 | **31.276** |
|  | SSIM | 0.7705 | 0.9732 | 0.9762 | **0.9978** |
| ROI-3 | RMSE($10^{-4}$) | 48.00 | 26.09 | 9.754 | **6.719** |
|  | PSNR | 46.416 | 51.668 | 60.216 | **63.453** |
|  | SSIM | 0.3343 | 0.4071 | 0.7889 | **0.8972** |
| ROI-4 | RMSE($10^{-4}$) | 66.50 | 34.35 | 14.43 | **10.281** |
|  | PSNR | 43.544 | 49.281 | 56.816 | **59.761** |
|  | SSIM | 0.4192 | 0.6105 | 0.8710 | **0.9474** |
| ROI-5 | RMSE($10^{-4}$) | 87.83 | 62.22 | 38.06 | **33.640** |
|  | PSNR | 41.127 | 44.122 | 48.389 | **49.463** |
|  | SSIM | 0.4443 | 0.6114 | 0.8605 | **0.9275** |

In this work, all the involved methods are programmed with Matlab (version 2014b) on a PC (i7-6700, 8.0 GB memory). The proposed DLIMD algorithm mainly includes two subroutines: dictionary training and material decomposition. As for the computational cost with material decomposition, the higher the number of the materials is, the greater the computational cost is. Particularly, the DIWET, DI, TVMD and DLIMD consume 0.04, 22.79 24.16 and 44.17 seconds, respectively. Besides, the DLIMD needs more time (387.22 seconds) to train the dictionary compared with other regularization based methods. Obviously, the DLIMD algorithm requires more computational cost than other algorithms.

### B. Preclinical experiments

In this study, the PILATUS3 PCD with 4 energy-channels by DECTRIS is employed to acquire multi-energy projections in preclinical application. Such PCD consists of 515 cells and each has a length of 0.15 mm. Here, a full-scan is performed to obtain projections with 4 energy bins from 720 views. Fig. 7 (a) shows the imaged object consisting of chicken foot and 5 mg/mL iodine solution cylinder. The distances from x-ray source to rotational center and detector are 35.27 cm and 43.58 cm, respectively. The FBP reconstruction results (see Fig. 7 (b)-(e)) have $512 \times 512$ pixels and each of them covers an area of $0.122 \times 0.122 \text{mm}^2$.

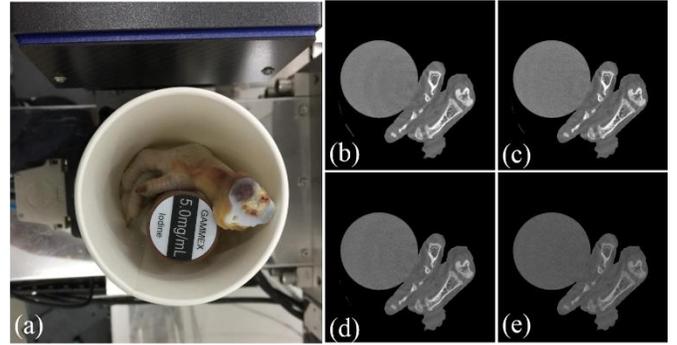

Fig. 7 Artificial mixed tissues experiment. (a) is the preclinical specimen fixed on the spectral CT system. (b)-(e) are FBP reconstruction results from 4 energy bins, where the display window is [0 0.5] cm$^{-1}$.

To evaluate the advantages of the DLIMD method, the material decomposition results from FBP results with different algorithms are shown in Fig. 8. For bony components, the image edge of bony structure is blurred in DI results. Although the TV based material decomposition method can provide clear image edge, the blocky artifacts are appeared in component map. This point can be confirmed by the bony structure indicated with red arrow within the magnified bony ROI "A" in Fig. 9. As for soft tissue decomposition results, the DLIMD can obtain clear image structure and features. To clarify this point, three soft tissue ROIs marked with "B", "C" and "D" are extracted and magnified in Fig. 9. From Fig. 9, it can be easily seen that image edge and structures are remarkably clear than those obtained by TVMD and DI methods. Similar to the soft tissue components, as for iodine contrast, the DI can wrongly classify the pixels of bone component into iodine contrast followed by TVMD results. Obviously, the DLIMD method can furtherly improve the accuracy of decomposed materials. This conclusion can be observed from the magnified ROI "E" in 3$^{rd}$ row of Fig. 8. According to Fig. 7(a), the iodine concentration is 5.0mg/mL and it can be considered as gold standard to evaluate the decomposition accuracy of different algorithms. Here, the RMSEs and mean value of extracted ROI 6 are calculated in Table III. From Table III, Here, the DLIMD can obtain the largest mean value of iodine and the smallest RMSE, followed by TVMD and DI methods with FBP reconstruction case.

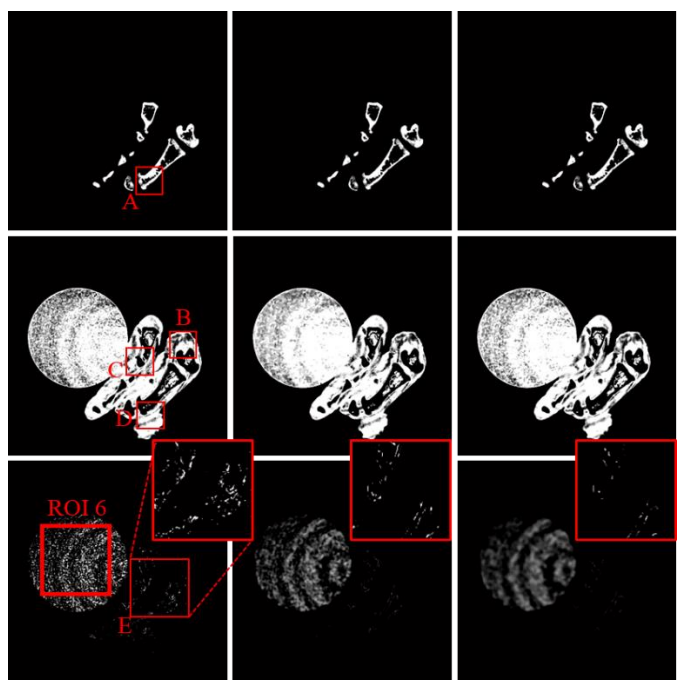

Fig. 8 Material decomposition results from Fig.7. From left to right, the columns represent the DI, TVMD and DLIMD methods. The 1$^{st}$-3$^{rd}$ rows represent the bone, soft tissue and iodine with the display windows [0.25 0.5], [0.85 0.95] and [0.0018 0.005], respectively.

The advanced spectral CT image reconstruction methods can improve the accuracy of material decomposition by providing high quality reconstructed images [25]. To further demonstrate the advantages of the proposed DLIMD algorithm in advanced reconstruction result case, the NLCTF reconstruction method is employed to implement image reconstruction and corresponding final material decomposition results using different algorithms are furtherly given in Fig. 10. Since the TVMD can obtain higher material decomposition accuracy than DI method, only the decomposition results from TVMD and NLCTF methods are given to save space in Fig. 10. Although the TVMD method can provide the higher quality of bone component map, the blocky artifacts still appear. This point can be confirmed by the extracted ROIs "F" and "G" which are further magnified in in Fig. 11. From 1$^{st}$-2$^{nd}$ columns of Fig. 11, it can be observed that the image structure from TVMD results indicated by "3" contains blocky artifacts. Fortunately, this structure can be observed in DLIMD results. Besides, the gap between two bones indicated by arrow "4" is still blurred. However, it is clear in the proposed DLIMD method. As for soft tissue decomposition results, the DLIMD can still obtain higher image quality with clear image edges and much feature. Specifically, the image structure is blurred in TVMD in Fig. 10, but DLIMD can provide clear image edge. Especially, two ROI labelled by "H" and "K" are extracted and magnified in Figs. 10 and 11. The image features indicated by arrows "2" and "6" are disappeared in TVMD. They can still be seen in DLIMD results. In addition, the image structure is too blurring to observe shape of edge. However, we can see the details of edge in DLIMD results. Here, the MSEs and mean value from location of ROI-6 are also calculated in Table III. From Table III, one can see the DLIMD can obtain the smallest MSE. In terms of mean value, the TVMD and DLIMD can obtain the same accuracy.

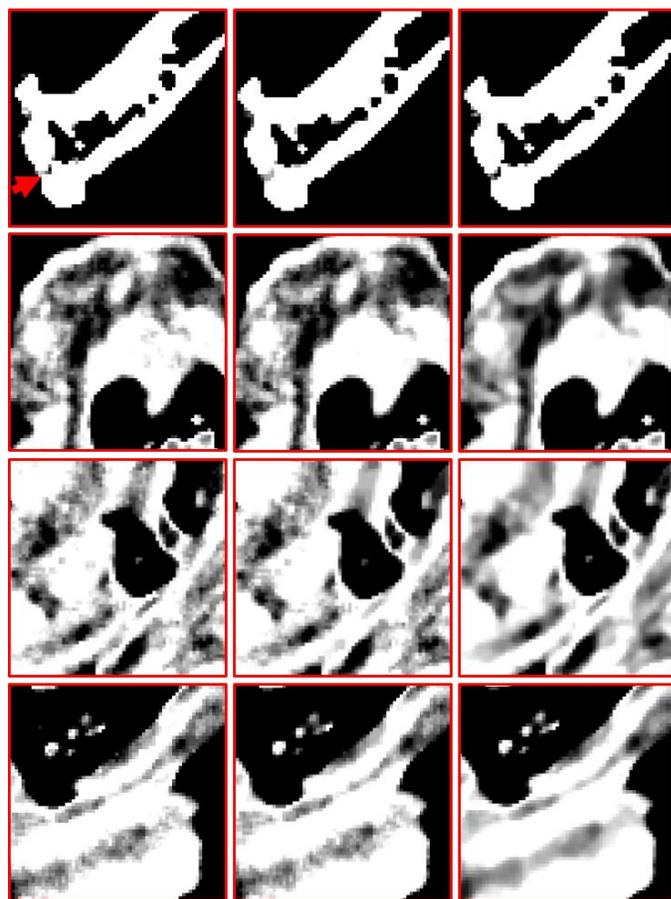

Fig. 9 The magnified ROIs from Fig.8. From left to right, the columns represent the DI, TVMD and DLIMD methods. The 1$^{st}$-4$^{th}$ rows represent the ROIs marked with "A", "B", "C" and "D", where the display windows are [0.29 0.33], [0.85 0.95], [0.85 0.95] and [0.85 0.95], respectively.

Table III. Quantitative evaluation results of ROI-6.

|  |  | DI | TVMD | DLIMD |
|---|---|---|---|---|
| FBP | Mean (%) | 0.1864 | 0.2282 | **0.2283** |
|  | MSE ($10^{-6}$) | 11.62 | 7.750 | **7.629** |
| NLCTF | Mean (%) | 0.1830 | **0.2353** | 0.2353 |

| | MSE (10⁻⁶) | 10.36 | 7.245 | **7.234** |

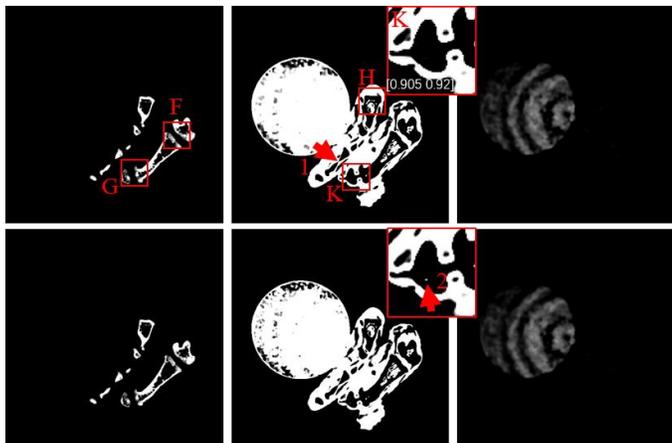

Fig. 10 Material decomposition results from the NLCTF. From left to right, the columns represent the bone, soft tissue and iodine with the display windows [0.25 0.5], [0.90 0.92] and [0.0018 0.005], respectively. The 1st-2nd rows represent TVMD and DLIMD methods.

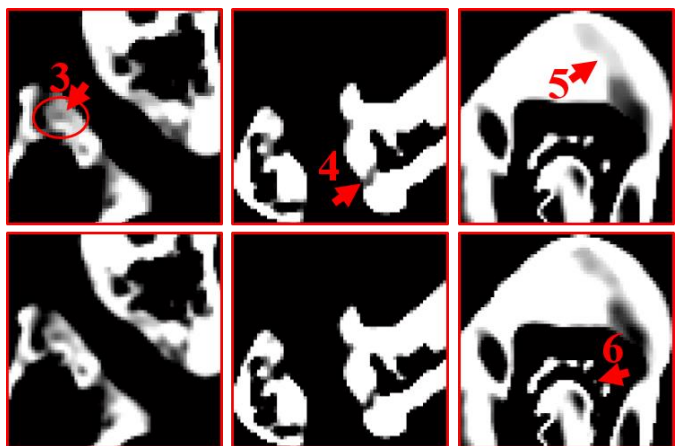

Fig. 11 The magnified ROIs in Fig.10. 1st-2nd rows represent the TVMD and DLIMD methods. The 1st-3rd columns represent the ROIs marked with "F", "G" and "H", where the display windows are [0.3 0.45], [0.29 0.33] and [0.89 0.94], respectively.

## V. DISCUSSIONS AND CONCLUSIONS

To improve the accuracy of material decomposition for spectral CT, we propose a DLIMD technique. Compared with the previous image-domain material decomposition method for spectral CT [12], the innovations of DLIMD are threefold. First, considering the similarities of different material images, we construct a unified dictionary to code material image sparsity by training a set of image patches. Here, those image patches are extracted from normalized material images. Second, we formulate a DLIMD mathematical model by enhancing sparsity of material maps with the dictionary and analyzing the image-domain material decomposition. Third, additional constraints (volume conservation [7] and the bound of each material pixel value) are incorporated into basis material decomposition model to further improve the decomposition accuracy. Two real datasets from physical phantom and preclinical dataset are employed to evaluate the proposed method, and the results demonstrate the advantages of DLIMD in improving image quality and material decomposition accuracy.

As a material decomposition method in image domain, the DLIMD can remove the artifacts to some extent or improve the accuracy of material images from x-ray beam hardening. For example, although aluminum component in the employed physical phantom can cause FBP image quality reduction and contain x-ray beam hardening artifacts. However, the DI, TVMD and DLIMD methods with enhanced constraints can still provide high accuracy of material decomposition results.

Because the reconstructed spectral CT images usually contain noise and artifacts, they can cause the instabilities of image domain material decomposition. A feasible strategy is to formulate regularization-based decomposition models. While TV is a common regularizer to be chosen, we introduce an advanced sparsity representation (i.e., dictionary learning) into the material decomposition model for better performances in terms of image feature recovery and edge protection. Although the DLIMD method can obtain satisfied decomposition results in image-domain, there are still some remaining issues that should be addressed. First, the parameters are only empirically optimized by comparing extensive experiment results. For example, regarding the selection of $\varepsilon$, it is necessary to select an $\varepsilon_m$ for each material component independently. However, if $m$ is great, it is difficult to select appropriate $\varepsilon$. In this case, it is feasible to select a reasonable $\varepsilon$ by estimating noise level within all material images in our follow-up work. Besides, we will also try some automatic strategies to optimize other parameters in the DLIMD method. Second, two real datasets only contain three different basis materials. However, the imaging objects may contain multiple (greater than 3) materials. It is necessary to validate and evaluate the performance of DLIMD in such cases.

In conclusion, based on the dictionary learning theory, we propose a DLIMD method for image-domain material decomposition of spectral CT. Real dataset experiments demonstrate the outperformance of DLIMD technique. This will be extremely useful for image domain material decomposition in practical applications of spectral CT.